\documentstyle[epsfig]{mn}
\def\be{\begin{equation}}
\def\ee{\end{equation}}
\def\bea{\begin{eqnarray}}
\def\eea{\end{eqnarray}}
\def\etal{{et al.}\thinspace}

\begin{document}

\title
[Entropy of the intracluster medium at high redshift]
{Entropy of the intracluster medium at high redshift}

\author[Biman B. Nath]
{Biman B. Nath\\
Raman Research Institute, Bangalore 560080, India\\
(biman@rri.res.in)
}
\maketitle

\begin{abstract}
{
Recent observations of entropy of intracluster medium (ICM) at high
redshift are not consistent with a redshift independent entropy floor.
In order to interpret these observations, we study models in which the entropy
of ICM is enhanced at the epoch of cluster formation ($z_f$),
which then passively
evolves as a result of radiative cooling until the epoch of observation ($z_o$).
We confirm that a $z$-independent entropy floor is incompatible with
the observations.
We find that models in which energy deposition into the ICM increases
with redshift (scaling approximately 
as $\sim M_{cl}^{0.6} (1+z_f)^{2.2}$) are consistent with observations
of low and high redshift clusters. Possible sources of such non-gravitational
heating are briefly discussed.
}
\end{abstract}

\begin{keywords} Cosmology: Theory---Galaxies: Intergalactic Medium---
Galaxies : clusters : general---
X-rays: Galaxies: Clusters
\end{keywords}

\section{Introduction}
X-ray observations of ICM in galaxy clusters have shown its entropy
to be larger than that expected from pure gravitational processes (Ponman 
\etal 2003, and references therein).
These observations have led to the suggestion that the entropy of ICM has been 
enhanced. Possible mechanisms for this include `preheating' the gas
 before its infall into the cluster potential (Kaiser 1991), shock heating 
 during the accretion
into the cluster (Tozzi \& Norman 2001; Babul \etal 2002; Voit \etal
2003), 
radiative cooling (Voit \& Bryan 2001), 
and energy input from supernova
driven winds (Wu \etal 2000) or active galaxies (Velageas \& Silk 1999;
Nath \& Roychowdhury 2002). 
Recent observations of lack of an entropy
core in the ICM in low mass clusters have been however difficult to interpret
in terms of an entropy floor that is expected in `preheating' scenarios
(Ponman \etal 2003; Pratt \& Arnaud 2003).

Very recently, observations of the ICM entropy 
 at high redshift have shown that a $z$-independent
entropy floor is difficult to reconcile with observations (Ettori \etal 2004).
The observed entropy 
(defined as $S\equiv T/n_e^{2/3}$) is found to
scale with redshift as $SE_z^{4/3} \propto (1+z)^{0.3}$ (where 
$E_z=H(z)/H_0$ is the ratio of Hubble constants at redshifts $z$ and zero).
In the $\Lambda$CDM universe with $\Omega_0=0.3$ and $\Omega_\Lambda=0.7$,
$E_z^{4/3} \propto (1+z)^{0.9}$, which would have led to a stronger
scaling of entropy ($SE_z^{4/3}$) with redshift than observed if entropy
is determined by a $z$-independent floor.

It is important
to compare these observations with expectations
from simple considerations of redshift scaling of various parameters.
For example, in the absence of entropy enhancement due to any of 
the processes
mentioned earlier, one expects the clusters forming at high redshift
to be denser than their low redshift counterparts, thereby making the
entropy lower than at low redshift for a given cluster mass. 
It is not clear how the requirement
of extra entropy should scale with redshift, and this is what we propose
to discuss in the present paper.

One important point with regard to the redshift scaling is that these
scalings
refer to the redshift of observation of a cluster, $z_o$, whereas
any physical model to understand the scaling must involve the formation
redshift ($z_f \ge z_o$) of the cluster, as well as
 the evolution thereafter until $z_o$
(e.g., Kitayama \& Suto 1996).
%In the standard structure formation theory, $z_f \ge z_o$.
%To understand the scaling with redshift, it is important to distinguish
%between the redshift of observation $z_o$ and the redshift of formation,
%$z_f$, of a cluster (e.g, Kitayama \& Suto 1996).
In the hierarchical structure formation theory, haloes merge to form bigger
haloes as time elapses. Although haloes also become bigger by slow accretion
of matter, merging between haloes is the dominant mechanism of driving haloes
upward in the hierarchy of structure. In this scenario, there are epochs
in between major mergers during which haloes evolve slowly by
accretion at a low rate, during which the gravitational potential does
not change substantially. In other words, when a cluster is observed
at $z_o$ it is reasonable to assume that it has been evolving with
a reasonably stable gravitational potential since an earlier epoch,
which can be termed its formation redshift $z_f \ge z_o$.  
Observational scalings of entropy with $z_o$ must therefore be translated
to requirements of excess entropy which is likely to depend on the formation
redshift $z_f$, and the duration of evolution between $z_f$ and $z_o$.

In this Letter, we build models of galaxy clusters formed at high
redshift, based on simple extrapolation of those used for clusters
formed at the present epoch. We then study the passive evolution of
ICM as a result of radiative cooling between the epochs of formation 
observation. We enhance the entropy of ICM in various ways at its
formation redshift and compare with the observations. We then discuss
the plausible models of entropy enhancement which are consistent with
observations.

Throughout the paper we use the $\Lambda$CDM
cosmological model with $\Omega_\Lambda=0.7$,
$\Omega_m=0.3$ and $h=0.7$.

\section{ICM at high redshift}
We study the distribution of intracluster gas assuming that the total gas
mass is very small compared to the total mass of the cluster, and that
the gravitational field is dominated by the dark matter. We first determine
a `default profile' that the ICM is expected to achieve, in the absence
of any non-gravitational modifications to its entropy (\S 2.1). 

\subsection{Gas profile without entropy enhancement}
We assume the ICM to be initially
 in hydrostatic equilibrium with the gravitational
potential of the dark matter. The dark matter is assumed to follow the
universal Navarro-Frenk-White (NFW) profile, $\rho_{dm}=\rho_s /((r/r_s) (1+
(r/r_s)))$, where $r_s=r_{vir}/c$, `c' being the concentration parameter and
$r_{vir}$ being the virial radius (Navarro \etal 1997). 
The virial radius is fixed by the
overdensity estimated from spherical collapse model, for the appropriate
redshift of the formation of the cluster $a_f$. 
The overdensity is approximately
given in a $\Lambda$CDM universe
by $\Delta (z)= 18 \pi ^2 + 82  x - 39 x^2 \,$, where  $x=\Omega_m(z_f)-1$ 
(Bryan \& Norman 1998), and whose value at $z=0$ is approximately $\sim 100$
($\Omega_m(z=0)$ being assumed to be $0.3$). The characteristic density
$\rho_s$ for a cluster with virial mass $M_{cl}$ is then given by,
\be
\rho_{\rm s} = c^3 {M_{cl}
\over 4 \pi r_{\rm vir}^3} \Bigl [{\ln(1+c) - {c \over
(1+c)}\Big ]} ^{-1},
\label{eq:rhos}
\ee
We use the results of the simulations by Bullock \etal (2001) to estimate
the concentration parameter as,
\be
c=9\Bigl ({M_{cl} \over 1.5 \times 10^{13}h^{-1}
M_{\rm \odot}}\Bigr )^{-0.13} \, (1+z_f)^{-1}\,.
\label{eq:cpfit}
\ee
%so that low mass clusters at a given redshift are more concentrated than
%rich clusters, and clusters of the same mass which formed 
%at higher redshift are less
%concentrated than those formed at present.

To facilitate the comparison of our results with observations which are
often expressed in terms of the radius $R_{200}$ where the overdensity
is $200$, we compute this radius and present our results in its terms.

We assume, in accordance with the results from numerical simulations
by Loken \etal (2002) (see also Borgani \etal (2004); Yepes \etal 2004), 
that gravitational interaction
of baryons with the dark matter gives rise to a `universal temperature
profile', which is approximated by $T(r)=1.33 T_{ew} \,
(1+1.5 r/r_{vir})^{-1.6}$, where $T_{ew}$ is the emission
weighted temperature of the cluster. We estimate the emission weighted
temperature from the empirical mass-temperature relation for local clusters
and include a redshift dependent factor in this relation that is expected
from theoretical considerations. The empirical relation found by
Finoguenov \etal (2001) is given by, 
\be
M_{\rm  500}=3.57\times 10^{13}
\Bigl ( {k_b
T_{ew} \over 1 \, {\rm keV}} \Bigr )^{1.58} 
{\rm M}_{\odot}
\label{eq:M-T}
\ee
where $M_{500}$ is the mass contained within the radius where the overdensity
is $500$. This relation is valid for clusters with $M_{500}
\ge 5 \times 10^{13}$ M$_{\odot}$ and
is a reasonable assumption for the default profile for all clusters.

We have used the temperature
 profile of Loken \etal (2002) in our calculations because
of the convenient analytical fit. It is possible that profiles obtained
by other adiabatic 
simulations (e.g., Borgani \etal (2004)) are somewhat different
than the one used here. A temperature profile which is shallower in the
inner region than that used
here may lead to a slight overestimation of the required excess entropy
in the final result (for details of the implications of the `universal
temperature profile', see Roychowdhury \& Nath 2003). 

The expected scaling of the mass-temperature relation 
with redshift $z_f$ of formation of clusters is
$T \propto E_{z_f}^{2/3} M^{2/3}$,
%clusters is $M \propto E_{z_f}^{-1} T^{3/2}$, 
where $E_{z_f}=H(z_f)/H(z=0)=
[\Omega_0 (1+z_f)^3+\Lambda_0]^{1/2}$ (e.g, Afshordi \& Cen 2002). 
We therefore multiply the observed
mass-temperature relation (equation \ref{eq:M-T}) 
with a factor of $E_z^{-1}$ on the right hand side to determine the
default profile at the formation redshift.

The initial gas density profile is then assumed to be determined by the
hydrostatic equilibrium condition, ${dp \over dr}=- \rho_{g} GM(r)/r^2$,
where the temperature profile is determined as explained earlier. We normalize
the gas density distribution in such a way that the total gas mass within
the radius $R_{200}$ is a constant fraction $f_{g,200}=
0.133$ of the total cluster mass,
as recently found by Ettori (2003) for high and low redshift clusters. 

We express our results in terms of the emission-weighted temperature
of the ICM, calculated with the help of the Raymond-Smith code for
a metallicity of $Z=0.3 Z_{\odot}$.

\subsection{Enhancement of entropy}
We assume two simple methods of entropy enhancement to the initial gas
profile. Let us define $\sigma \equiv P/\rho ^{\gamma}$ as the
`entropy index', where $\gamma$ is the adiabatic index. The entropy
per unit particle can be written in terms of this entropy index as
$s={\rm constant}+({\gamma -1}) k_B \ln (\sigma)$, 
where $k_B$ is the Boltzmann constant. In terms of this
entropy index, we assume that in the first type of entropy enhancement,
\be
\sigma (r)=\max (\sigma_0 , \sigma_i (r) ) \,,
\label{eq:en_in1}
\ee
where $\sigma_i(r)$ refers to the initial gas profile and $\sigma_0$ is
a constant, for all radii and for all clusters. Physically, this
corresponds to an entropy floor which may be caused by preheating. 
Below we will refer to this model as the `preheating' case.

In the second type of entropy enhancement, we assume that,
\be
\sigma (r)=\eta \, \sigma _i (r) \,,
\label{eq:en_in2}
\ee
where $\eta$ is assumed to be a
constant for the given cluster, for simplicity. Physically this
type of entropy enhancement
would correspond to energy deposition into the ICM after its infall
within the cluster potential well. The excess energy per particle
thus deposited, $\Delta \epsilon = T \Delta s \propto T
{\Delta \sigma \over \sigma} \propto T (\eta-1)$. Since gas temperature
is a weak function of radius, this roughly corresponds to a near uniform
deposition of energy throughout the cluster, for a constant value 
of $\eta$. We will refer to this as the `post-infall heating' case
in the following.

\subsection{Halo formation epoch}
Since haloes increase their mass by mergers and steady accretion, it is
difficult to assign a single epoch of formation for a given mass. 
Lacey \& Cole (1993)
 defined the halo formation epoch $z_f$ for a halo
of mass $M$ to be the epoch at which the halo has a mass greater than $M/2$
for the first time. After this epoch (`of formation') the halo is supposed
to attain the total mass $M$ by steady accretion of small objects. 
%In other
%words, this particular epoch signifies the epoch of last major merger 
%experienced by the observed halo. 
Lacey \& Cole (1993)
 then determined the 
distribution function for the probability that a halo of mass
$M$ that is observed at $z$ was formed at $z_f$, or had a mass greater than
$M/2$ for the first time at $z_f$. 

For our calculations, we define the halo formation epoch as the epoch
when the halo has a mass greater than ${3 \over 4} M$ for the first
time. This is motivated by the results of numerical simulations
which show that gravitational potential does not change much after
$75\%$ of the total mass is in place (Navarro \etal 1995; see also Babul \etal
2002).

\begin{figure}
\centerline{
\epsfxsize=0.4\textwidth
\epsfbox{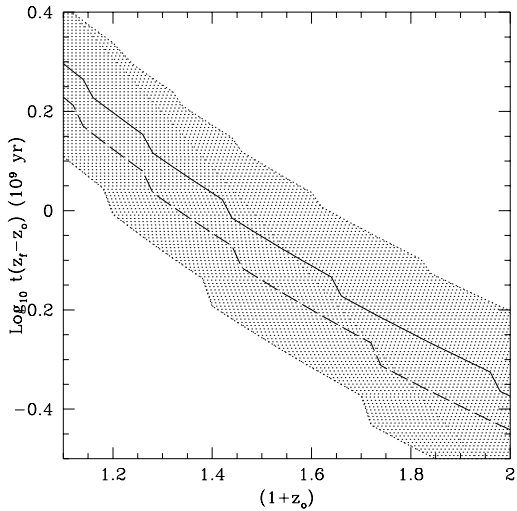}
}
%{\vskip-3mm}
\caption{
The duration of evolution, between the redshift of formation, $z_f$ and
that of observation $z_o$ of clusters of mass $10^{14}$ M$_{\odot}$ is
shown against $1+z_o$. The thick solid line corresponds to the median
value of distribution of $z_f$ and the shaded region corresponds to
$30\%$ and $70\%$ values of $z_f$. The dashed line corresponds to
the median value of $z_f$ for clusters of mass $10^{15}$ M$_{\odot}$.
}
\label{f:age}
\end{figure}

For our calculations, we use the power spectrum for $\Lambda$CDM
cosmology %using the transfer function of Bardeen \etal (1986) and
normalized by $\sigma_8=0.9$, and for growth of linear perturbation
in this cosmology,
we use the fit by Lahav \etal (1991).. 
Using the probability distribution we can determine 
different percentiles of the formation epoch $z_f$. 
We show in Figure \ref{f:age} the expected time of evolution, {\it i.e.}
the time spent between the formation redshift $z_f$ and when it is observed
at $z_o$, as a function of $z_o$ for clusters of $10^{14}$ M$_{\odot}$.
The solid line represents the median value of the distribution in $z_f$
and the shaded region the $30\%$ and $70\%$ percentile values. The
dashed line shows the median value for $z_f$ distribution for clusters
of mass $10^{15}$ M$_{\odot}$. It is clear that evolution is more
important for clusters observed at lower redshifts. 

We find that a convenient fit for this duration is
given by (for the median value) $\Delta t \approx 2.5 \times 10^9 {\rm yr} \,
(1+z_o)^{-2.6} (M_{cl}/10^{14}
M_{\odot})^{-0.09}$ which is accurate within $5\%$ for
$z_o \le 1$.

\section{Evolution of gas in clusters}
We divide the gaseous sphere (which resides within the potential
well provided by the dark matter, $M_{dm}\approx M_T$, which is the total 
matter) into mass shells which are designated
by gas mass contained within them, $M_g(r)$. 
In terms of the entropy index introduced earlier, the hydrostatic
equilibrium condition  reads,
\bea
{dP \over dM_g}=&& -{G M_T (r) \over 4 \pi r^4}
\nonumber\\
{dr \over dM_g}=&& {1 \over 4 \pi r^2} (P/\sigma)^{-1/\gamma}\,,
\eea
The default profile is computed using these equations and the universal
temperature profile. We set the  boundary condition 
 that the gas pressure at the mass shell which was at 
a radius $R_{200}$ for the default profile, and for which
$M_g/M_{cl}=f_{g,200}=0.133$, remains a constant. 

The entropy of the gas is then enhanced
instantaneously (according to equations \ref{eq:en_in1} and
\ref{eq:en_in2}). This enhancement changes the density and
temperature profile of the gas (pushing it outward to some extent; see below).

We assume that after the initial (impulsive) enhancement,
intracluster gas cools radiatively and evolves in a quasi-hydrostatic
manner.
For subsequent evolution as a result of radiative cooling, if
$\Lambda (t)$ denotes the cooling function, the specific entropy
evolves with time as (see also Kaiser \& Binney 2003),
\be
{ds \over dt}={1 \over \gamma -1} k_B {1 \over \sigma} {d \sigma \over dt}=
-{\Lambda (T) \over T} n\,,
\label{eq:cool}
\ee
where the electron density $n_e=0.85 \rho (r) /m_p$.
We use the cooling function appropriate
for metallicity $Z=0.3 \, Z_{\odot}$ from Sutherland \& Dopita
(1993) for which Nath (2002) provided a fit. The equation \ref{eq:cool}
is integrated for each mass shell for a small time interval $\delta t$
and the entropy index added to this shell after the time interval is,
\be
\Delta \sigma =-(\gamma -1) \sigma n {\Lambda (T) \over k_B T} \delta t \,.
\ee
The equations of hydrostatic equilibrium are used again to determine the
density and temperature profile at time $t+\delta t$.

\section{Results}
\begin{figure}
\centerline{
\epsfxsize=0.4\textwidth
\epsfbox{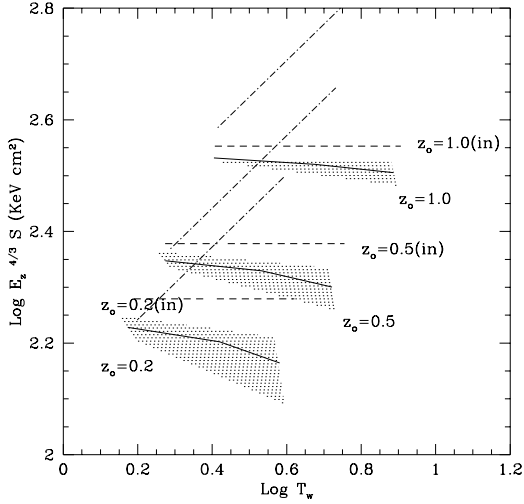}
}
{\vskip-3mm}
\caption{
Plot of entropy (at $0.1 R_{200}$) for
clusters with different $T_{ew}$ and observed at redshifts $z_o=0.,
0.5, 1.0$.
The thick solid lines in each redshift set
 correspond to the median value of distribution
of $z_f$ and the shaded region (with dots) to the range between
$30\%$ and $70\%$ values of $z_f$. For comparison, the initial
entropy of the same clusters are shown with dashed lines (and
labeled `in').
The three slanted dotted lines
show the observed scaling of entropy-temperature relation, $E_z^{4/3}S
\propto T^{0.65} (1+z_o)^{0.3}$, corresponding to the three different
redshifts $z_o=0.,
0.5, 1.0$.
}
\label{f:floor}
\end{figure}

We first show the results for the case of preheating, {\it i.e.} the case
of an entropy floor (equation \ref{eq:en_in1}), in Figure \ref{f:floor}.
It shows the
observed entropy at $0.1 R_{200}$ for clusters with different
emission weighted temperatures and observed at $z_o=0.0,0.5,1.0$
for an assumed value of $\sigma_0=0.27 \times 10^{34}$ erg cm$^2$ g$^{-5/3}$.
(For comparison, Tozzi \& Norman (2001) used a value of $0.3 \times 10^{34}$
erg cm$^2$ g$^{-5/3}$ for their calculations.) This corresponds to 
$S\equiv T/n_e^{2/3} \approx 247.9$ keV cm$^2$ (assuming $n_e \approx 
0.875 \rho /m_p$).
It is constructed out of data points calculated for clusters of different
masses observed at different epochs. The extreme left points
in each set correspond to clusters of $5 \times 10^{13}$ M$_{\odot}$
and the extreme right points use clusters with $5 \times 10^{14}$ M$_{\odot}$.
The thick solid line uses the median value of distribution of $z_f$,
and the shaded region corresponds to the range between $30\%$ and
$70\%$ values of $z_f$.

\begin{figure}
\centerline{
\epsfxsize=0.4\textwidth
\epsfbox{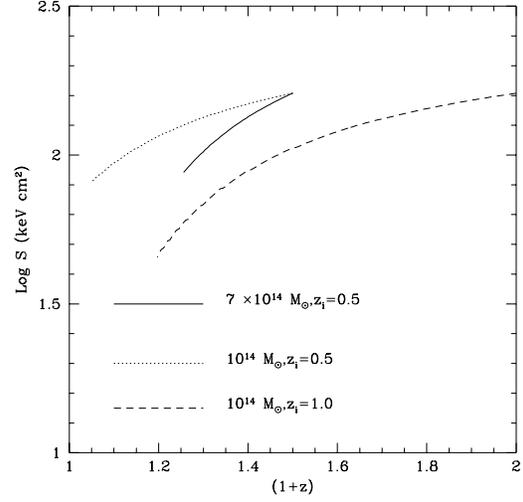}
}
{\vskip-3mm}
\caption{
Evolution of entropy at $0.1 R_{200}$ for clusters of different masses
with different redshifts of formation is shown.
Dotted and dashed lines
correspond to clusters of mass $10^{14}$ M$_{\odot}$ with formation
redshift $0.5$ and $1.0$ respectively, and solid line correspond to,
$5 \times 10^{14}$ M$_{\odot}$ with $z_f=0.5$.
}
\label{f:track}
\end{figure}

Figure \ref{f:floor} leads to a few important conclusions. Firstly, it is clear
that a $z$-independent entropy floor is inconsistent with both the
observed S-T relation at low redshift and its scaling with redshift. The
S-T relation that follows from it is in general shallower than the
observed $S \propto T^{0.65}$ scaling at the present epoch (Ponman \etal 2003). 
This result has been earlier noted by Borgani \etal (2001) from numerical
simulations.
The evolution of ICM between
the formation redshift and that of observation makes this problem
worse. This is because gas in massive clusters (with higher $T_{ew}$)
loses entropy at a faster rate. This is elaborated in Figure \ref{f:track},
which shows a few examples of evolutionary tracks of entropy with
redshift for different clusters with various $z_f$. The solid and
the dotted lines show the evolution of a massive and poor cluster,
respectively, with the same $z_f$. Clearly the ICM in massive
cluster loses entropy faster than the ICM in low mass clusters, making
the observed S-T relation in Figure \ref{f:floor}
 flatter than the initial S-T relation
at formation. Moreover, since the duration of evolution (between
$z_f$ and $z_o$ is longer at low redshifts, the flattening of S-T
relation is prominent at lower redshifts.

Secondly, the evolution of the S-T relation with redshift is stronger
than observed (shown with dotted lines). This has already been pointed
out by Ettori \etal (2004). Our calculations show that the problem is
made worse than previously thought as a result of evolution of ICM
between $z_f$ and $z_o$.

We then turn to the case of post-infall heating. We have found that
assuming $\eta$ to be independent of cluster mass and redshift produces
a steeper S-T relation than observed, and a scaling with redshift slower
than observed.
%results inconsistent with both the observed S-T relation and its
%scaling with redshift. 
We therefore used simple power-law scalings for $\eta$
with cluster mass and redshift in the form, $\eta (M_{cl},z_f)
\propto M_{cl}^{\alpha} (1+z_f)^\beta$, and determined the approximate
values of $\alpha$ and $\beta$ that produce  results consistent with
observations.

\begin{figure}
\centerline{
\epsfxsize=0.4\textwidth
\epsfbox{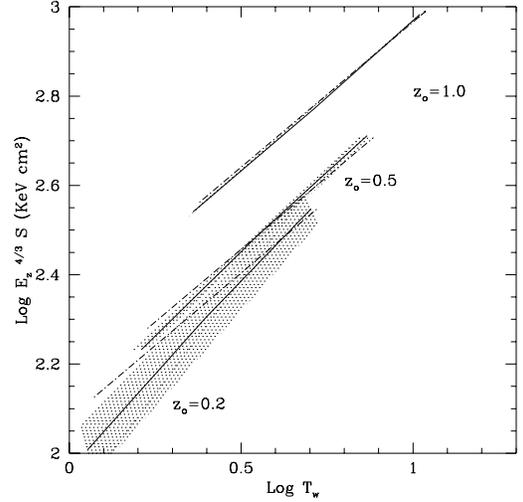}
}
{\vskip-3mm}
\caption{
Same as in Figure \ref{f:floor} but for the case of `post-infall'
entropy enhancement, with $\eta \propto M_{cl}^{-0.3} (1+z_f)^{2.2}$.
}
\label{f:dep}
\end{figure}

We show in Figure \ref{f:dep} the results for the `post-infall' heating
case, with $\eta =10 \, (M_{cl}/5 \times 10^{13} M_{\odot})
^{-0.3} (1+z_f)^{2.2}$, along with the
observed S-T relation and its scaling with redshift (shown with dotted
lines). We find that
this form of dependence of $\eta$ with cluster mass and formation
redshift is reasonably consistent with observations.

\begin{figure}
\centerline{
\epsfxsize=0.4\textwidth
\epsfbox{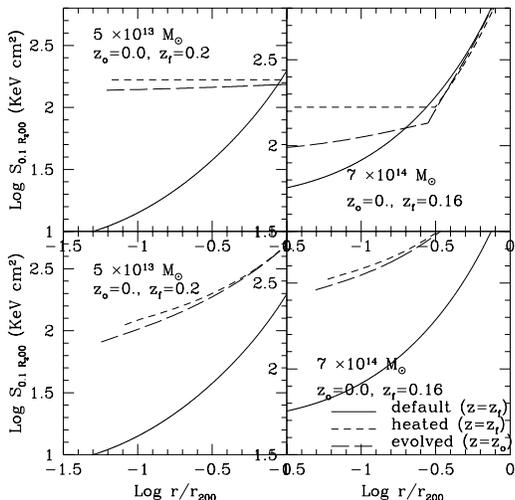}
}
{\vskip-3mm}
\caption{
Entropy ($S=T/n_e^{2/3}$) profiles are shown at different epochs for
two types of entropy injection and for two cluster masses. The left
panels represent clusters of mass $5 \times 10^{13}$ M$_{\odot}$
with $z_f=0.2$ and the right panels, $7 \times 10^{14}$ M$_{\odot}$
$z_f=0.16$. The upper panels show the case of $z$-independent entropy
floor (equation \ref{eq:en_in1}) and the lower panels show the
case of `post-infall' heating
(equation \ref{eq:en_in2}). 
The solid curves show the initial profile, the dashed lines show
the profile immediately after entropy injection and the long-dashed
curves show the result of evolution until the redshift of observation
at $z_o=0.$.
}
\label{f:ex}
\end{figure}

Finally, we show a few examples of the entropy profiles, before and after
the entropy enhancement and after the passive evolution until the epoch
of observations, for a couple of clusters with varying masses in
Figure \ref{f:ex}.
The solid line shows
the initial entropy ($S\equiv T/n_e^{2/3}$) profile
at formation epoch. The dashed curve then shows the profile after entropy
enhancement. The upper panels show the result of using equation (
\ref{eq:en_in1}) for entropy injection and the lower panels use equation 
(\ref{eq:en_in2}). As expected, the intracluster gas is pushed out
as a result of this entropy injection. Finally the long dashed curves
show the result of evolution until the redshift of observation ($z_o=0$
in this case). The left panels show the case of $M=5 \times 10^{13}$
M$_{\odot}$ and the right panels that of  $M=7 \times 10^{14}$ M$_{\odot}$.
The redshift of formation for these two cases have been chosen to be
the corresponding median values of distribution of $z_f$ ($0.16$
and $0.20$).

%\begin{figure}
%\centerline{
%\epsfxsize=0.4\textwidth
%\epsfbox{fig6.ps}
%}
%{\vskip-3mm}
%\caption{
%Evolution of entropy at $0.1 R_{200}$ (multiplied by the factor $E_z^{4/3}$)
% is shown against the redshift
%of observation, for clusters of masses $5 \times 10^{13}$ (upper set of
%curves) and 
%$5 \times 10^{14}$ M$_{\odot}$ (lower set). The scalings $(1+z)
%)^{0.3}$ and $(1=z_o)^{0.6}$ are shown for comparison with dotted lines.
%}
%\end{figure}

%\begin{figure}
%\centerline{
%\epsfxsize=0.4\textwidth
%\epsfbox{fig7.ps}
%}
%{\vskip-3mm}
%\caption{
%The observed entropy ($E_z^{4/3} S$) at $0.1 R_{200}$ for clusters
%of masses $ 10^{14}$ and $5 \times 10^{14}$ M$_{\odot}$ is shown
% as a function of the emission weighted temperature $T_{ew}$ for
%clusters observed at different redshifts. The middle set of points
%correspond to the median value of distribution in $z_f$ and the
%outer dots, to $30\%$ and $70\%$ values. The left most dots in
%each set correspond to $z_o=0.2$ and the next one on the right is 
%separated by $\Delta z_o=0.1$.
%}
%\end{figure}

%\begin{figure}
%\centerline{
%\epsfxsize=0.4\textwidth
%\epsfbox{fig8.ps}
%}
%{\vskip-3mm}
%\caption{
%Plot of entropy (at $0.1 R_{200}$) that would be observed for
%clusters with different $T_{ew}$ at different redshifts $z_o=0.2,
%0.5, 1.0$.
%The thick solid line corresponds to the median value of distribution
%of $z_f$ and the shaded region (with dots) to the range between
%$30\%$ and $70\%$ values of $z_f$. For ready comparison, the initial
%entropy of the same clusters are shown as regions shaded
%with dashed lines (and marked with `in').
%}
%\end{figure}

\section{Discussion}
We discuss the implication of the result arrived at the last section
that energy deposition in ICM with $\eta \propto M_{cl}^{-0.3} (1+z_f)^{2.2}$
is consistent with observations of low and high-$z$ clusters. Firstly, since
the excess energy deposited per particle is proportional to $T (\eta-1)
\sim T \eta$ (since $\eta > 1$),
our assumption of a constant $\eta$ for a given cluster roughly amounts  
to uniform deposition of energy throughout the cluster, since temperature
is a weak function of radius (it decreases by much less than an order of
magnitude from the core to the outermost radius for the default profile).
Secondly, since $T \propto M_{cl}^{2/3}$,
the variation of $\eta$ with $M_{cl}$
implies that the required energy per particle $T \eta \sim 10 \,
{\rm keV} M_{cl}^{1/3}
$.
Since gas mass is proportional to the total mass, the above
result implies that the total required amount of energy $\Delta E
\propto \Delta \epsilon M_{cl} \propto M_{cl}^{1.3}$. 

The required scaling of $\eta$ with redshift ($\propto (1+z_f)^{2.2}$)
is mainly due to the fact that densities are larger at higher redshift,
%outcome of densities being larger at higher redshift,
reducing the default entropy level, and thereby increasing
the requirement. 
This increase is somewhat compensated by the relatively short
duration of evolution (between $z_f$ and $z_o$) at high redshift (
see Figure \ref{f:age}) but not significantly. 

It is of course possible that the enhancement of entropy is more
complicated than represented by
 equations \ref{eq:en_in1} and \ref{eq:en_in2}. Especially
the form of energy deposition into the ICM after its infall can either
depend on the radius and/or time, in which case the required scaling of
$\eta$ would be different from the results presented here. A non-trivial
radial dependence of $\eta$ may change the entropy profile substantially
and change the form of passive evolution thereafter. For example, in the
case of a strong radial dependence, the entropy in the inner region may
increase substantially, with the possibility of a negative entropy gradient
ushering convective flows (e.g., Roychowdhury \etal 2004). 

Our assumption of an instantaneous entropy enhancement followed by
passive evolution has been admittedly simple, since
our goal has been to interpret the general dependence of observed entropy with
redshift. It is however possible that the period of entropy enhancement
is substantial (see, e.g., Roychowdhury \etal 2004), but modelling such
cases introduces another free parameter (of enhancement timescale)
in the calculation.
Although detailed modeling of such
complicated forms of heating may change
the scaling power index of $\eta$
with redshift to some extent, but is not expected to substantially
 change the strong
redshift dependence
of $\eta$, which mainly arises from densities being larger at high
redshift.

It is interesting to note that the final entropy profiles in the 
case of near-uniform  heating after infall 
(see lower panels of Figure \ref{f:ex})
are consistent with recent observations of positive
gradient of entropy even at small radii for poor clusters, instead
of a core as expected from the preheating scenarios (upper panels
of Figure \ref{f:ex}) (Ponman \etal 2003; Pratt \& Arnaud 2003). 
The near isentropic core in the case of an entropy floor is
consistent with the numerical simulations of Borgani \etal (2001)
(considering that the value of the entropy floor used here is larger
than assumed by them), and also Babul \etal 2002 (who used a value
of $\sigma_0$ that is $1.7$ times that used here).
We must however remember that a strong radial
dependence of heating  of ICM may change the profile by depositing
more energy (and entropy) in the central region as compared to the
outer region. In other words, the recent observations of positive
gradient of entropy at inner radii is consistent with a (post-infall)
 near-uniform energy
deposition into the ICM throughout the cluster.

As far as the plausible sources of heating which can explain the
observations of high redshift ICM, it is interesting to note
that rate of heating by active galaxies is expected to rise with
redshift (Nath \& Roychowdhury 2002) mainly because of the fact that
quasar activity has declined in the universe in the recent past.
The amount of energy deposited is also reasonably consistent
with estimates from heating from buoyant bubbles from active
galaxies (Roychowdhury \etal 2004).
%decrease with cluster mass in this scenario (Nath \& Roychowdhury 2002). %Also,
%Begelman (2004) has argued that the amount of energy deposited into
%the ICM may scale linearly with the cluster mass, which is close
%to the above mentioned requirement of $\Delta E \propto M_{cl}^{1.3}$.
In this regard, one should also consider the possibility of supernova
driven winds as the source of heating, since the total energy deposited
in this case is also expected to increase with the cluster mass (total
mass of galaxies, from which the winds would emanate, being proportional
to the total cluster mass). Moreover since the average star formation rate
in the universe
increases with redshift ( for $z \le 1$), it is possible to achieve
a strong redshift dependence in this case. We however note that
detailed calculations have shown supernova driven winds to be inefficient
sources of heating (Wu \etal 2000).

\section{Summary}
We have calculated the expected evolution of entropy  of ICM (at $0.1 R_{200}$)
with redshift (of observation $z_o$), taking into account passive
evolution of ICM between the epochs of formation and observation of the
cluster. We found that
the adoption of a $z$-independent entropy floor exacerbates the problem
of accounting for the observed evolution of the entropy-temperature
relation with redshift.
If the entropy
is enhanced instead with the help of a heating source after the infall of the
ICM, we find that the total energy required to be deposited scales
as $\Delta E \propto M_{cl}^{1.3} (1+z_f)^{2.2}$.

\bigskip
It is a pleasure to thank S. Roychowdhury and S. Sridhar for stimulating
discussions, and the anonymous referee for valuable suggestions.

\end{document}